\documentclass[12pt]{article}
\usepackage{epsf}
\usepackage{latexsym}
\usepackage{a4}
\newcommand{\Ref}[1]{(\ref{#1})}
\newcommand{\be}{\begin{equation}}
\newcommand{\ee}{\end{equation}}
\newcommand{\bea}{\begin{eqnarray}}
\newcommand{\eea}{\end{eqnarray}}
\newcommand{\beao}{\begin{eqnarray*}}
\newcommand{\eeao}{\end{eqnarray*}}
\newcommand{\nn}{\nonumber}
\newcommand{\pa}{\partial}
\newcommand{\e}{{\rm e}}
\renewcommand{\d}{{\rm d}}
\newcommand{\E}{{\cal E}}
\renewcommand{\P}{{\bf P}}
\newcommand{\om}{{\omega}}
\newcommand{\ep}{{\epsilon}}
\begin{document}
\title{Ground State Energy for Massive Fields and Renormalisation}

\author{M Bordag \\
        \small Institute for Theoretical Physics, University of
               Leipzig,\\ 
        \small  Augustusplatz 10, 04109 Leipzig, Germany\\
        {\small file is \jobname}}

\maketitle

\begin{abstract}
  We discuss the renormalisation of the ground state energy of massive fields
  obeying boundary conditions, i.e., of the Casimir effect, and emphasise the
  role of the mass for its understanding. This is an extended abstract of a
  talk given at the topical group meeting on Casimir Forces at the
  Harvard-Smithsonian Center for Astrophysics on March 15-29, 1998.
\end{abstract}
Renormalisation is a key to the understanding of the structure of quantum
field theory.  The kind of ultraviolet divergencies occurring divide the
perturbative field theoretical models into renormalizable and non or super
renormalizable ones. The ultraviolet divergencies occurring during the
calculation of ground state energy in different backgrounds (including
boundary conditions, i.e., the Casimir effect) carry information on the
classical system which one is forced to associate in order to remove resp.
interprete the divergencies.

In the present note which is an extended abstract of a talk\footnote{given at
  the topical group meeting on Casimir Forces at the Harvard-Smithsonian
  Center for Astrophysics on March 15-29, 1998.}  we discuss the
renormalisation using different examples. These are the Casimir effect for
massive scalar \cite{ms} and spinor \cite{sp} fields and the radiative
corrections \cite{rk} to the electromagnetic Casimir effect with boundary
conditions on a sphere. In addition we consider the ground state energy of a
scalar field in a spherically symmetric smooth background field \cite{gz3d}.
For all questions and references not given in this note we refer to the cited
papers. 

The necessity to associate some classical system with any ground state energy
arises from its very nature. The ground state energy is the amount of energy
left in a quantised system when all excitations are gone. To any excitated
level it adds the same amount of energy (or other observables like charge)
and, hence, cannot be observed in any measurements involving a mere change of
the quantum numbers (for example in transitions between different levels).
The only way to observe ground state energy is to change parameters which are
external to the quantised system. In the most prominent example, the Casimir
effect between conducting plates, this external parameter is the distance
between the plates. Therefor it is inevitable to introduce an appropriate
classical system.

The consideration of parallel plates is to some extend misleading. The point
is that it is a too simple example, hiding most of the classical structures
which one is forced to introduce in the general case. From a very general
point of view it is the missing curvature in the case of parallel plates which
makes the interesting divergent contributions vanish. What is left is the so
called Minkowski space contribution which is independent on the distance
between the plates so that one obtains a finite force without any further
renormalisation.  As a result of this simple behaviour there have been
attempts to extend this to the sphere using for example two concentric spheres
in order to extract a finite force. Also, efforts have been undertaken to use
the remarkable properties of the Zeta functions in order to get infinity free
results. As we will argue in the present note, this cannot be successful in
the general case.

\vspace{2cm}

\noindent
We consider the following models\footnote{We  use units where $\hbar=c=1$.}:

\begin{enumerate}
\item a 'smooth'\footnote{The word 'smooth' means here the opposite of
    boundary conditions. In fact, the only example which we consider up to the
    end is a square well potential which is not smooth in the mathematical
    sense.} background field.  The action is given by \bea S&=&\frac 1 2
  \int\d x~\left\{\Phi(x) (\Box -M^2 -\lambda \Phi (x)^2 )\Phi(x)
    +\nn\right.\\
  &&\left. ~~~~~~~~~~~~~~\varphi(x) (\Box -m^2 -\lambda ' \Phi(x)^2
    )\varphi(x)\right\},\eea where $\Phi(x)$ is the classical background field
  (we choose it static and spherically symmetric) and $\varphi(x)$ is to be
  quantised. The complete energy of that system is
  \bea \E&\equiv&\E_{\rm class}+\E_{\rm qu}\label{Egesbar}\\
  &= &\frac{1}{2} \int \d \vec{x} ~\left(\left(\nabla
      \Phi(x)\right)^{2}+M^{2}\Phi(x)^{2}+\lambda \Phi(x)^{4}\right)+\frac 1 2
  \sum\limits_{(n)}\left(\lambda_{(n)}^{2}+m^{2}\right)^{\frac 1 2},\nn\eea
  where the one particle energies are defined by the eigenvalue problem \be
  (-\Delta +\lambda'\Phi(\vec{x})^{2})\varphi_{(n)}(\vec{x})=\lambda_{(n)}^{2}
  \varphi_{(n)}(\vec{x}).\label{EV}\ee

\item a scalar field with boundary conditions on a sphere
\be (\Box -m^{2})\varphi (x)=0 , \qquad \varphi(x)_{|_{r=R}}=0. \ee

\item a spinor field in a bag
\be (i\gamma^{\mu}{\pa\over \pa x^{\mu}}+m)\psi(x)=0, \qquad (1+i\vec{n}
\vec{\gamma})\psi(x)_{|_{r=R}}=0.\ee
\end{enumerate}
The last two examples can be subdivided with respect to the region where we
consider the quantum field:\\[5pt]
\begin{tabular}{rrcll}
$i)$   & $0\le r\le R$ &    &                 &interior,\\
$ii)$  &               &    & $R\le r< \infty$&exterior,\\
$iii)$ & $0\le r\le R$ &$\cup$& $R\le r< \infty$&both regions.
\end{tabular}\\[5pt]
The third case, $iii)$, looks like the simple union of $i)$ and $ii)$.
It is,  in fact, with respect to the spectrum of the quantum system. 
Below we will see
how after  renormalisation a difference occurs. 

The ground state energy is given 
in all cases  by the formula 
\[\E_{\rm qu}=\pm\frac 1 2
\sum\limits_{(n)}\left(\lambda_{(n)}^{2}+m^{2}\right)^{\frac 1 2},\]
where the sum runs over the spectrum of the corresponding quantum field.  
The sign in front distinguishes between bosonic and fermionic
fields. In
2.$i)$ we have $(n)=(n,l,m_{l})$ and $\lambda_{(n)}=j_{\nu,n}/R$ with the roots of
the Bessel functions, $J_{\nu}(j_{\nu,n})=0$ ($\nu\equiv l+\frac 1 2 $). In
3.$i)$ the spectrum is determined by the equation
\[ J_{\nu+1}(\om R)+\sqrt{{E-m\over E+m}}J_{\nu}(\om R)=0~~\to 
~~\om=\lambda_{(n)}\]
with $E=\sqrt{m^{2}+\om^{2}}$.
For $ii)$ we have to consider some large but finite volume first in order to
get a discrete spectrum. By a well known procedure which is explained in
detail in \cite{ms} one lets that volume tend to infinity throwing away the so
called Minkowski space contribution.

In the last two examples the associated classical (geometrical) system
has the energy \be \E_{\rm class}=pV+\sigma S+FR+k+\frac h R,
\label{Eclass}\ee where $V=\frac 4 3 \pi R^{3}$ is the volume and $S=4
\pi R^{2}$ is the surface of the sphere. Correspondingly, $p$ is the pressure
and $\sigma$ is the surface tension. The parameters $F$, $k$ and $h$ do not
have a special meaning.  This formula is the most general one which can be
written down for dimensional reasons. It turns out that this form is required
in the cases $i)$ and $ii)$, while in $iii)$ the first, the second and the
third terms can be dropped, see below.

\vspace{2cm}

\noindent
The expression for the ground state energy written so far is divergent and
must be regularised. Because of its technical advantages (and its beauty) we
use the zeta functional regularisation and write \be \E_{\rm
  qu}=\pm\frac{\mu^{2s}}{2} \sum\limits_{(n)}(\lambda_{(n)}^{2}+m^{2})^{\frac
  1 2 -s}
\label{E0reg}\ee with sufficiently large $\Re s$ to make the sum
converge. In the end one has to perform the analytic continuation to $s=0$. In
Eq. \Ref{E0reg} the parameter $\mu$ with dimension of a mass was introduced.
It is arbitrary and similar to the subtraction point in the renormalisation of
perturbative quantum field theory.  After renormalisation the ground state
energy will become independent of $\mu$.

Eq. \Ref{E0reg} by means of
\be \E_{\rm qu}=\pm\frac{\mu^{2s}}{2}\zeta_{\P} (s-\frac 1 2 )\label{E0zeta}\ee
is in fact the expression of the ground state energy using the
zeta function of a corresponding operator $\P$
\be \zeta_{\P} (s)=\sum\limits_{(n)}e_{(n)}^{-s},\label{zeta}\ee
where $e_{(n)}=\sqrt{\lambda_{(n)}^{2}+m^{2}}$ are the eigenvalues: 
$\P\varphi_{(n)}=e_{(n)}\varphi_{(n)}$.

Later on it will be instructive to have another regularisation. 
We use
\be \E_{\rm qu}=\pm\frac{1}{2} \sum\limits_{(n)}(\lambda_{(n)}^{2}+m^{2})^{\frac 1
  2 } \e^{-\ep(\lambda_{(n)}^{2}+m^{2})}\label{E0regep}\ee
with $\ep\to 0$ in the end. 

The regularised ground state energy \Ref{E0reg} can be written in the form
(see the cited papers for details, for the dropping of the Minkowski space
contribution for instance) \be \E_{\rm qu} =-{\cos \pi s\over \pi}\mu^{2s}
\sum_{l=0}^{\infty}(l+1/2) \int\limits_{m}^{\infty}dk\,\,
[k^2-m^2]^{\frac{1}{2}-s}~\frac{\partial}{\partial k}\ln f_l
(ik).\label{E0d}\ee Here, all information is contained in the function
$f_{l}(k)$. In the first model it is the Jost function of the scattering
problem corresponding to \Ref{EV}, in the second and third models it is
expressed by the corresponding combinations of Bessel and Hankel functions:\\[6pt]
\begin{tabular}{ll}
2. $i)$: \quad &$f_{l}(k)\to J_{\nu}(kR)$,\\[4pt]
2. $ii)$: \quad &$f_{l}(k)\to H^{(1)}_{\nu}(kR)$,\\[4pt]
3. $i)$: \quad &$f_{l}(k)\to  J_{\nu+1}(k R)+\sqrt{{E-m\over E+m}}J_{
\nu}(k R)$,\\[4pt]
3. $ii)$: \quad &$f_{l}(k)\to  H^{(1)}_{\nu+1}(k R)+\sqrt{{E-m
\over E+m}}H^{(1)}_{\nu}(k R)$.
\end{tabular}

Thereby the integration runs over the imaginary axis starting from $im$
(cf. Eq. \Ref{E0d}). Representation \Ref{E0d}
is equivalent to other ones where  the mode
density or the scattering phase shift enter. The integration over the
imaginary axis has some technical advantages. For instance, there are no
oscillations in $f_{l}(ik)$ and the possible bound states which may occur in a
background potential like $\lambda'\Phi(\vec{x})^{2}$ in \Ref{EV} are included implicitly.

\vspace{2cm}

\noindent
The general structure of the ultraviolet divergencies can be obtained from the
heat kernel expansion. For this reason one represents the zeta function in
\Ref{zeta} by an integral \be \zeta_{\P}(s)=\int\limits_{0}^{\infty}{\d t
  ~t^{s-1}\over \Gamma (s)}~K(t),\label{zetap}\ee where \be
K(t)=\sum\limits_{(n)}\e^{-t(\lambda_{(n)}^{2}+m^{2})}\ee is the (global) heat
kernel. Now the ultraviolet divergencies of the ground state energy are
determined from the behaviour of the integrand in \Ref{zetap} at the lower
integration limit and, hence, from the asymptotic expansion of the heat kernel
for $t\to 0$: \be K(t)\sim {\e^{-tm^{2}}\over (4\pi
  t)^{3/2}}\sum\limits_{n}a_{n}t^{n}\qquad n=0,\frac 1 2 ,1,\dots
.\label{hke}\ee This expansion is known for very general manifolds, see the
book \cite{gilkey} for example. If the underlying manifold is without
boundary, only coefficients with integer numbers enter, otherwise half integer
powers of $t$ are present.  Sometimes logarithmic contributions occur (not
shown in \Ref{hke}), but we were not confronted with them so far.

Inserting the expansion \Ref{hke} into Eq. \Ref{zetap} we obtain from the
coefficients with $n\le 2$ (the higher coefficients do not contribute to the
ultraviolet divergencies)
\bea \E^{div}  &=& -\frac{m^4}{64\pi^2}\left(\frac 1 {s} 
+\ln\frac
{4\mu^2}{m^2} -\frac 1 2 \right) a_0 -{m^{3}\over 24\pi^{3/2}}a_{1/2}\nn\\ 
& &+\frac{m^2}{32\pi^2}\left(\frac 1 {s} 
+\ln\frac{4\mu^2}{m^2} -1\right) a_1+{m\over 16 \pi^{3/2}}a_{3/2}
\label{E0div}\\ 
& &-\frac 1 {32\pi^2} \left(
\frac 1 {s} +\ln \frac{4\mu^2}{m^2}-2\right) a_2. \nn\eea
Apparently, the contributions from the coefficients $a_{1/2}$ and $a_{3/2}$
are finite while the other contain a pole in $s=0$. This is a special feature
of the zeta functional regularisation
and not the case in other regularisations. So, for example, in the
regularisation given by Eq. \Ref{E0regep} the corresponding contribution
$\E^{\rm div}_{\rm ep}$ to
the ground state energy reads
\bea \E ^{\rm div}_{\rm ep}&=&{3\over 64 \pi^{5/2}}
\left\{\left(-\frac 4 3 {1\over \ep^{2}}+2 {m^{2}\over \ep}-{m^{4}\over
        2}\ln\ep\right)a_{0}+{\pi\over 2}\left({1\over \ep^{3/2}}-{m^{2}\over
      \ep^{1/2}}\right)a_{1/2}\nn\right.\\
&&\left.+\left({2\over \ep}+m^{2}\ln\ep\right)a_{1}+{\pi\over
    \ep^{1/2}}~a_{3/2}-\ln\ep ~a_{2}\right\}.\label{regep}\eea
Here the contributions of all coefficients are actually divergent. 

The coefficients in the first model are well known:
\be a_{0}={\cal V},~~a_{1}=-\int\d\vec{x}~\lambda'\Phi(\vec{x})^{2},~~a_{2}
=\frac 1 2
  \int\d\vec{x}~\lambda'^{2}\Phi(\vec{x})^{4}.\label{anbg}\ee
For the second model they read:
\be
\begin{array}{rclrclrcl} 
a_{0}&=&\pm {\cal V}, &a_{1/2}&=& -2\pi^{3/2}R^{2},\\[4pt]
a_{1}&=&\pm \frac 8 3 \pi R, &a_{3/2}&=& - {\pi^{3/2}\over 6}, 
&a_{2}&=&\mp {16\over 315}{\pi\over R},
\end{array}\label{anDir}\ee
where the upper sign corresponds to the model 2.$i)$, i.e., to the interior
and the lower sign to 2.$ii)$.  $\cal V$ is the volume of the underlying
manifold, the whole Minkowski space in the first model, the volume of the
interior of the sphere $\frac{4\pi}{3}R^{3}$ resp. the exterior volume in the
second model. In the first model the contribution of $a_{0}$ is independent of
the background potential and dropped. Similar arguments apply to 2.$ii)$.
 
The alternation of the signs is valid for general, infinitely thin bounding
surfaces, not for the sphere alone. When adding up to get $iii)$, the
corresponding ultraviolet divergencies cancel between inside and outside. This
is just the point, where the third case of the second and third models becomes
nontrivial as it has a smaller number of singular contributions than $i)$ and
$ii)$ taken individually.

\vspace{2cm}

\noindent
The renormalisation procedure consists simply in subtracting the divergent
contributions from the ground state energy and adding them to the classical
contribution: \bea \E&=&\underbrace{\E_{\rm class}+\E^{\rm div}_{\rm
    qu}}+\underbrace{\E_{\rm qu}-\E^{\rm div}_{\rm qu}}\nn\\
&\equiv&~~~~~~\tilde{\E}_{\rm class}~~~+~~~~~\E^{\rm ren}_{\rm qu}.
\label{renallg}\eea
The change from $\E_{\rm class}$ to $\tilde{\E}_{\rm class}$ can be
interpreted as a renormalisation of the parameters of the classical system. In
the first model it reads:
\bea M^{2}&\to&M^{2}-{\lambda'm^{2}\over 16\pi^{2}}\left(\frac 1 s
  +\ln{4\mu\over m}-1\right),\nn\\
\lambda&\to&\lambda-{\lambda'm^{2}\over 64\pi^{2}}\left(\frac 1 s
  +\ln{4\mu\over m}-2\right).\label{ren}\eea
The divergence associated with $a_{0}$ would lead to a renormalisation of a
 constant addendum to the classical energy. As said above, we drop such
 a contribution. We would only like to mention, that in a gravitational
background this  would be a renormalisation of the cosmological constant. 
It should be remarked, that the kinetic term in the classical action does not
undergo any renormalisation. 

In the second model, in the cases $i)$ and $ii)$, the procedure \Ref{renallg}
leads to the following substitutions:
\bea
p &\to &p \mp \frac{m^4} {64\pi^2}
\left( \frac 1 {s}
-\frac 1 2 +\ln \left[\frac{4\mu^2}{m^2}\right]\right),\quad
\sigma  \,\to  \, \sigma +\frac{m^3}{48\pi}, \nn\\
F  &\to &  F\pm \frac{m^2}{12\pi} 
\left( \frac 1 {s}
- 1  +\ln \left[\frac{4\mu^2}{m^2}\right]\right), \quad
k\, \to \, k-\frac m {96},
\label{n8}\\
h& \to & h\pm \frac 1 {630\pi}  
\left( \frac 1 {s}
- 2  +\ln \left[\frac{4\mu^2}{m^2}\right]\right).\nn
\eea
In contrast, for $iii)$ we need only
\be
\sigma  \to  \sigma +\frac{m^3}{24\pi}, \quad 
k\to k-\frac{m}{48}. \label{n8a}
\ee

So we conclude, that in $i)$ and $ii)$ all five contributions, introduced with
\Ref{Eclass}, are necessary in order to interpret the renormalisation as a
redefinition of the parameters of the classical system. In contrast, in the
case $iii)$, only two are required:
\be \E_{\rm class}^{iii)}=\sigma S+k.\label{Eclassiii}\ee 
Of course, more can be introduced depending
of the kind of physical systems  one has to consider. But this is the minimal
set. Sometimes the physical meaning (and real value) of these parameters is
unclear.  

The procedure of renormalisation is not unique. Besides different
regularisations which can be used and which change the details of formulas
like \Ref{n8} (for instance, when using \Ref{E0regep} and \Ref{regep} instead
of \Ref{E0reg} and \Ref{E0div}), there is an arbitrariness resulting from the
arbitrary parameter $\mu$ in \Ref{E0div}. Also, after the infinite
renormalisation \Ref{ren} or \Ref{n8}, a finite renormalisation of the same
kind is still possible. So one is asked to fix some normalisation condition in
order to give the ground state energy a unique meaning. As such we suggest
the requirement that the renormalised ground state energy must vanish when
making the quantum field infinitely heavy, i.e., for $m\to\infty$: \be \E^{\rm
  ren}_{\rm qu}\to 0 ~~~~\mbox{for}~~m\to\infty\label{normcond}.\ee This
condition is physically meaningful. In that limit there should be no quantum
fluctuations. On the other hand it is complete, i.e., it fixes $\E^{\rm
  ren}_{\rm qu}$ uniquely because the last coefficient contributing to
divergencies and, hence, the last which can cause a non-uniqueness is $a_{2}$
(in general $a_{d/2}$ where $d$ is the dimension of the manifold).  All lower
coefficients ($a_{n}$ with $n<\frac d 2$) enter proportional to nonnegative
powers of the mass.
 
The uniqueness of the ground state energy is essential when asking for
quantities like the Casimir force\footnote{Note that it is only for planar
  geometries where taking the derivative with respect to a distance removes
  the divergencies.}.
A different situations occurs when considering the back reaction
problem. There the dynamics of the background itself is included and one has
to look for a minimum of the complete energy $\E$ \Ref{Egesbar}
resp. \Ref{renallg} after renormalisation, varying the background (the field
$\Phi(\vec{x})$ or the radius of the sphere, for example). Obviously, in that 
case a finite renormalisation, i.e., adding zero by subtracting something
from $\E^{\rm div}_{\rm qu}$ and adding it to $\E_{\rm class}$, makes no
difference.

The goal of fulfilling the normalisation condition \Ref{normcond} is achieved
by subtracting the complete contribution resulting from the heat kernel
coefficients $a_{n}$ with $n\le 2$ as done by \Ref{ren} resp. \Ref{n8} and
\Ref{n8a} in the first resp.  second models.  In doing so we obtain $\E_{\rm
  qu}^{\rm ren}$ as defined by \Ref{renallg}. Now the regularisation can be
removed, i.e., the analytic continuation to $s=0$ can be performed. This is
still a nontrivial task because it cannot be done under the sign of the sum
and the integral in expressions like \Ref{E0d}. One has to use the uniform
asymptotic expansion of $\ln f_{l}(ik)$ for $l\to\infty$, $k\to\infty$ with
$\frac k l$ fixed to the required order ($l^{-3}$ in the considered examples).
This results in \be \E_{\rm qu}^{\rm ren}=\E_{f}+\E_{\rm as}-\E_{\rm qu}^{\rm
  div},\label{Equren}\ee where \be \E_{f}=-{1 \over \pi}
\sum_{l=0}^{\infty}(l+1/2) \int\limits_{m}^{\infty}dk\,\,
[k^2-m^2]^{\frac{1}{2}}~\frac{\partial}{\partial k}(\ln f_l (ik)-\ln f^{\rm
  as})\ee is the 'finite' part. Here, due to the achieved convergence one
could put $s=0$ under the sign of the integral. Also, it is possible to
integrate by parts yielding the representation \be \E_{f}={1 \over \pi}
\sum_{l=0}^{\infty}(l+1/2) \int\limits_{0}^{\infty}\d q ~(\ln f_l (ik)-\ln
f^{\rm as})_{|_{k=\sqrt{q^{2}+m^{2}}}}.\ee In the other ('asymptotic')
contribution, \be \E_{\rm as}=-{\cos \pi s\over \pi}\mu^{2s}
\sum_{l=0}^{\infty}(l+1/2) \int\limits_{m}^{\infty}dk\,\,
[k^2-m^2]^{\frac{1}{2}-s}~\frac{\partial}{\partial k}\ln f^{\rm
  as},\label{}\ee one has to perform the analytic continuation to $s=0$ which
is quite easy now because the structure of $\ln f^{\rm as}$ is much simpler
than that of $\ln f_{l}(ik)$. After that the divergences in $\E^{\rm as}$ must
cancel that of $\E^{\rm div}_{\rm qu}$ in \Ref{Equren}. In general, for $\ln
f^{\rm as}$ one can take the minimal asymptotic contributions as it was done
in the cited papers. But it is possible to include more (non-leading) terms,
for instance in order to speed up the convergence in $\E_{f}$. Once this
procedure is carried out (for details see the cited papers) the numerical
calculation of the ground state energy can be done.  

As an example we consider
here the result for the second model, i.e., for the scalar field with
Dirichlet boundary conditions on a sphere of radius $R$. For dimensional
reasons the result can be written as 
\be\label{29} \E^{\rm ren}_{\rm qu}={1\over R}f(Rm)
= m\, h(Rm),
\ee
 where $f(x)$ and $h(x)$ are dimensionless functions simply
connected by $f(x)=xh(x)$. In fact, the functions $f$ resp. $h$ show the
dependence of the energy on the mass $m$ resp. on the radius $R$.  The results
obtained in \cite{ms} for the function $f(x)$ are shown in figure 1.
\begin{figure}\unitlength=1cm
\begin{picture}(15,5)
\put(-0.5,0){\epsfxsize=5cm\epsfysize=5cm\epsffile{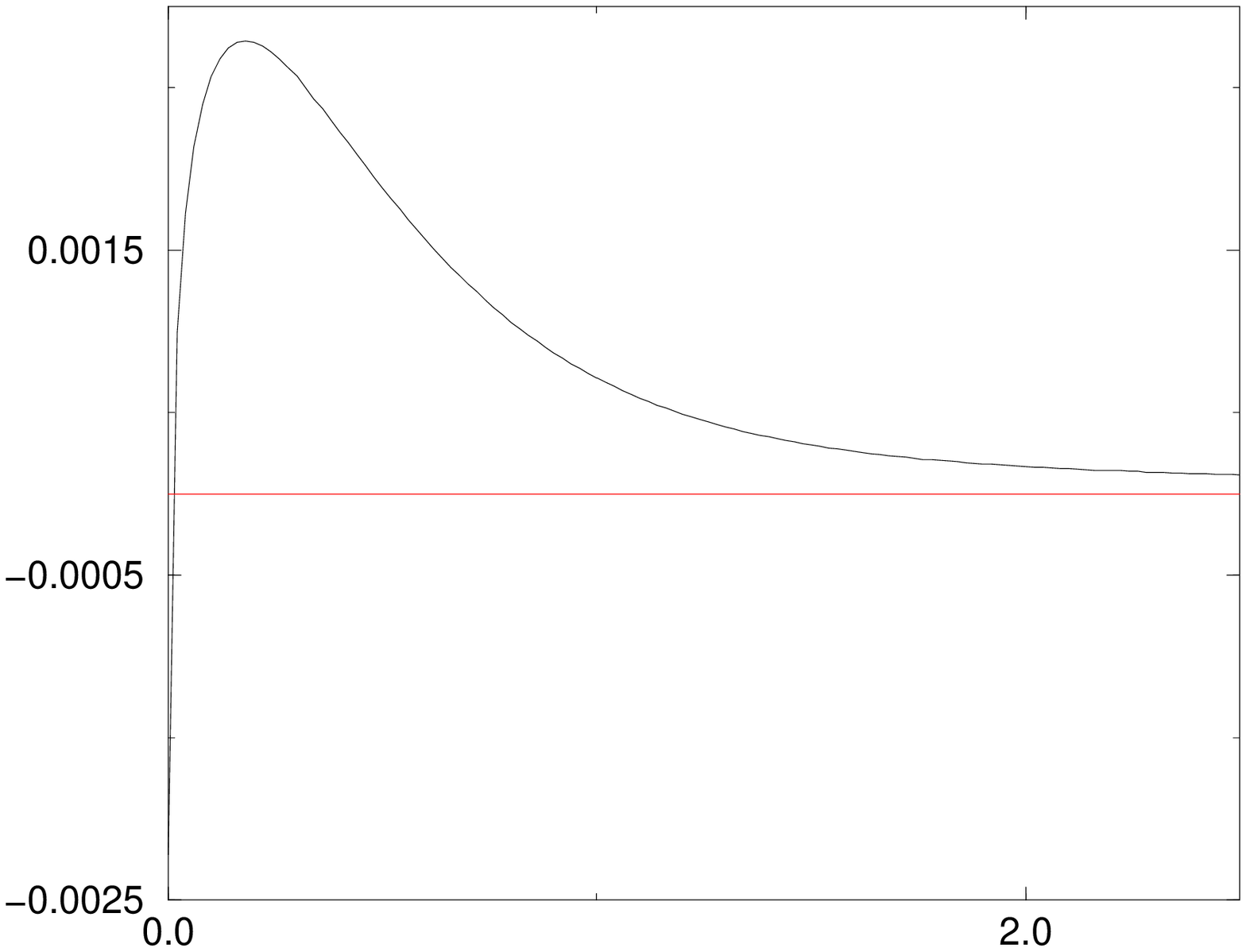}}
\put(4.5,0){\epsfxsize=5cm\epsfysize=5cm\epsffile{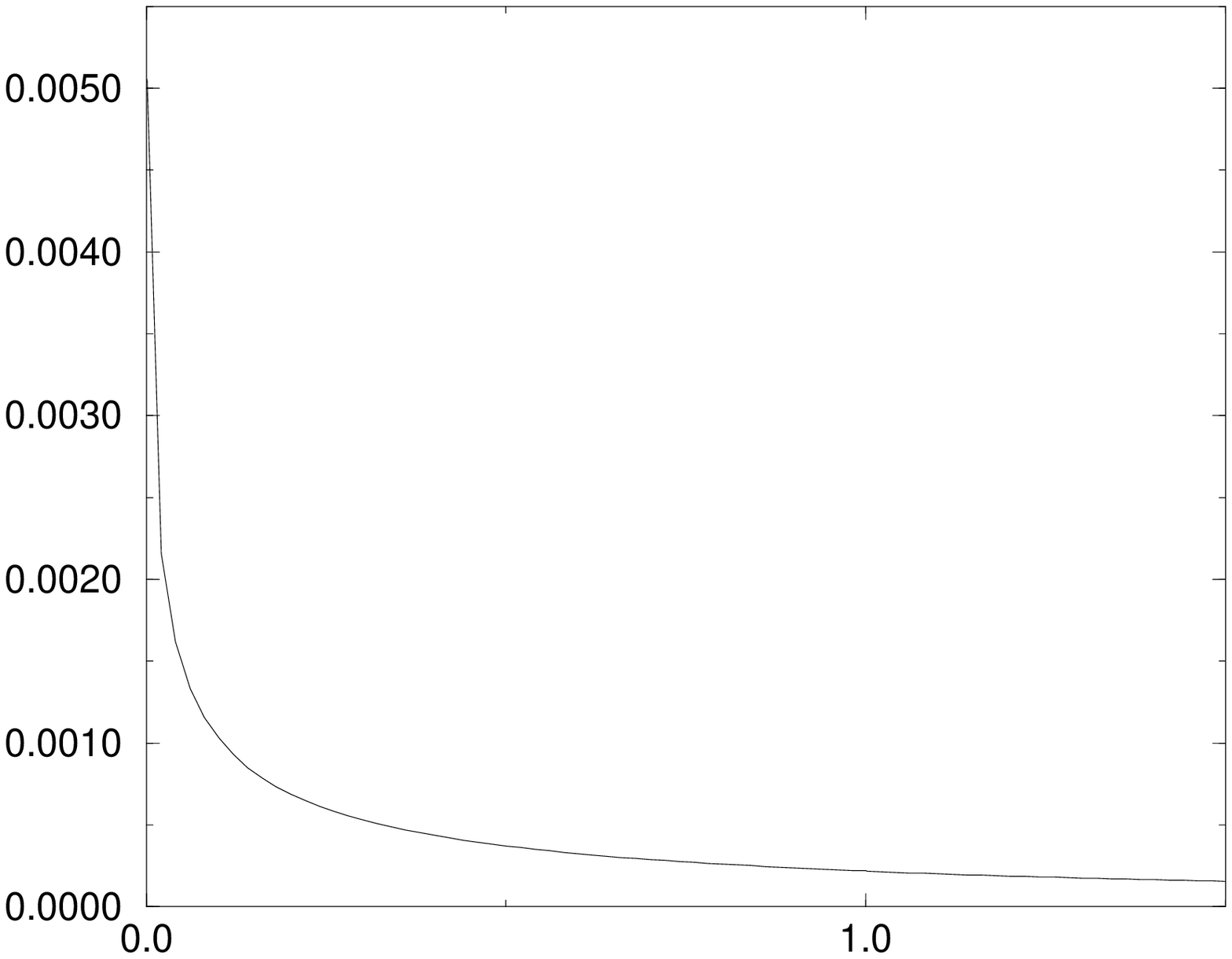}}
\put(9.5,0){\epsfxsize=5cm\epsfysize=5cm\epsffile{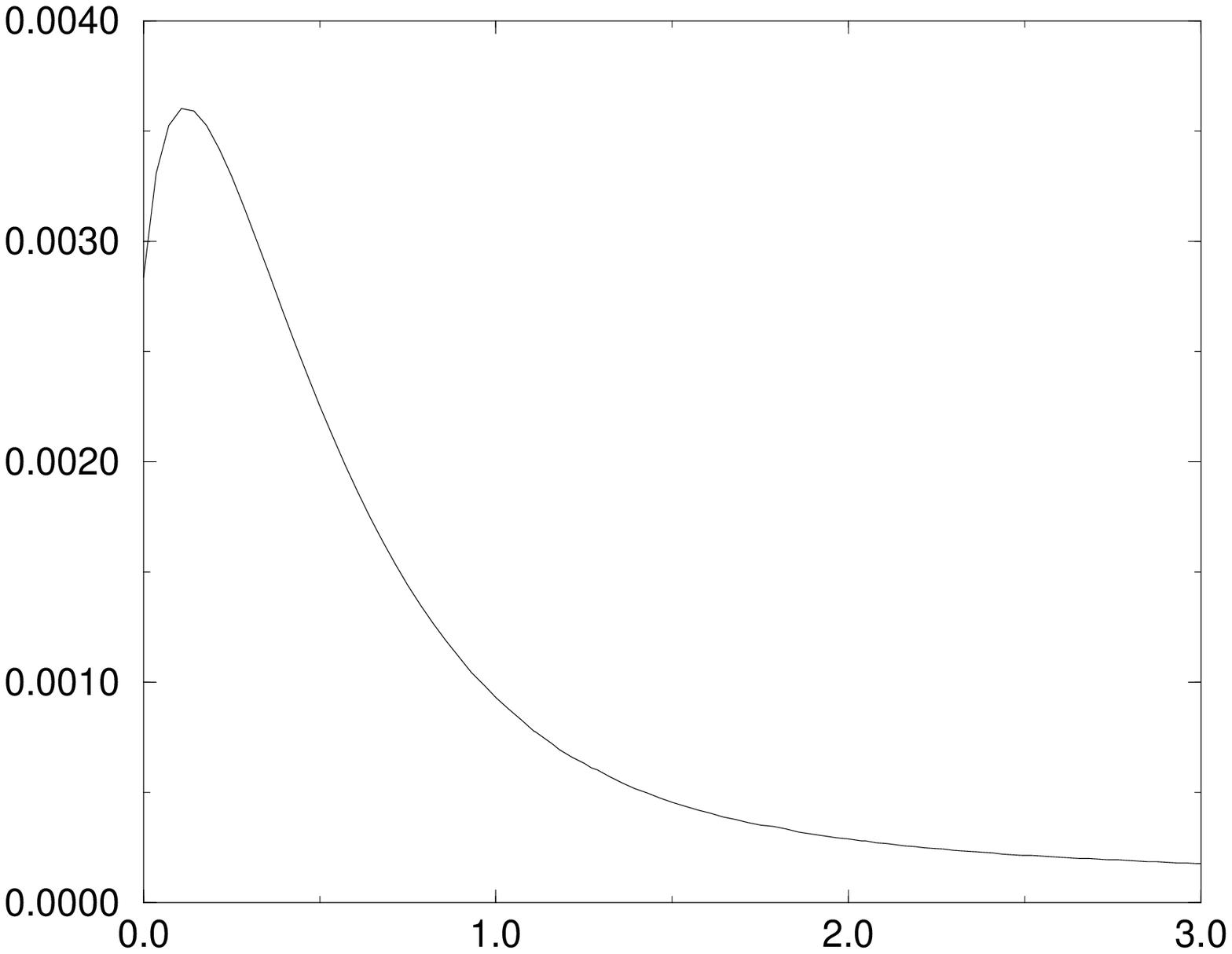}}
\put(14.2,0.6){$x$}
\put( 9.2,0.6){$x$}
\put( 4.2,0.6){$x$}
\put( 2.5,3.5){$i)$}
\put( 7.0,3.5){$ii)$}
\put(12.0,3.5){$iii)$}
\put(0,4.8){$_{f(x)}$}
\put(5,4.8){$_{f(x)}$}
\put(10,4.8){$_{f(x)}$}
\end{picture}
\caption{The functions $f(x)$ in case $i)$, $ii)$ and $iii)$ for the scalar 
field}
\label{abb1}
\end{figure}
It is interesting to note the maximums in $i)$ and $iii)$ for some finite
mass. In the function $h$, i.e., after dividing by $R$, and when passing to
the dependence on the radius, the minimum in case $i)$ survives. It is shown
in figure 2. In the other two cases, the dependence is simply monotonously
decreasing.

The same analysis as for the scalar case had been done in \cite{sp} for the
spinor field with bag boundary conditions. The result is almost the same, but
in detail different (for instance, with respect to the sign). As an example we
mention the dependence of the energy on the radius in the case $i)$, shown in
figure 3. Here, it is interesting to note the minimum.  

\begin{figure}[h]\unitlength=1cm
\begin{picture}(7.5,6)
\put(1,1){\epsfxsize=5cm\epsfysize=5cm\epsffile{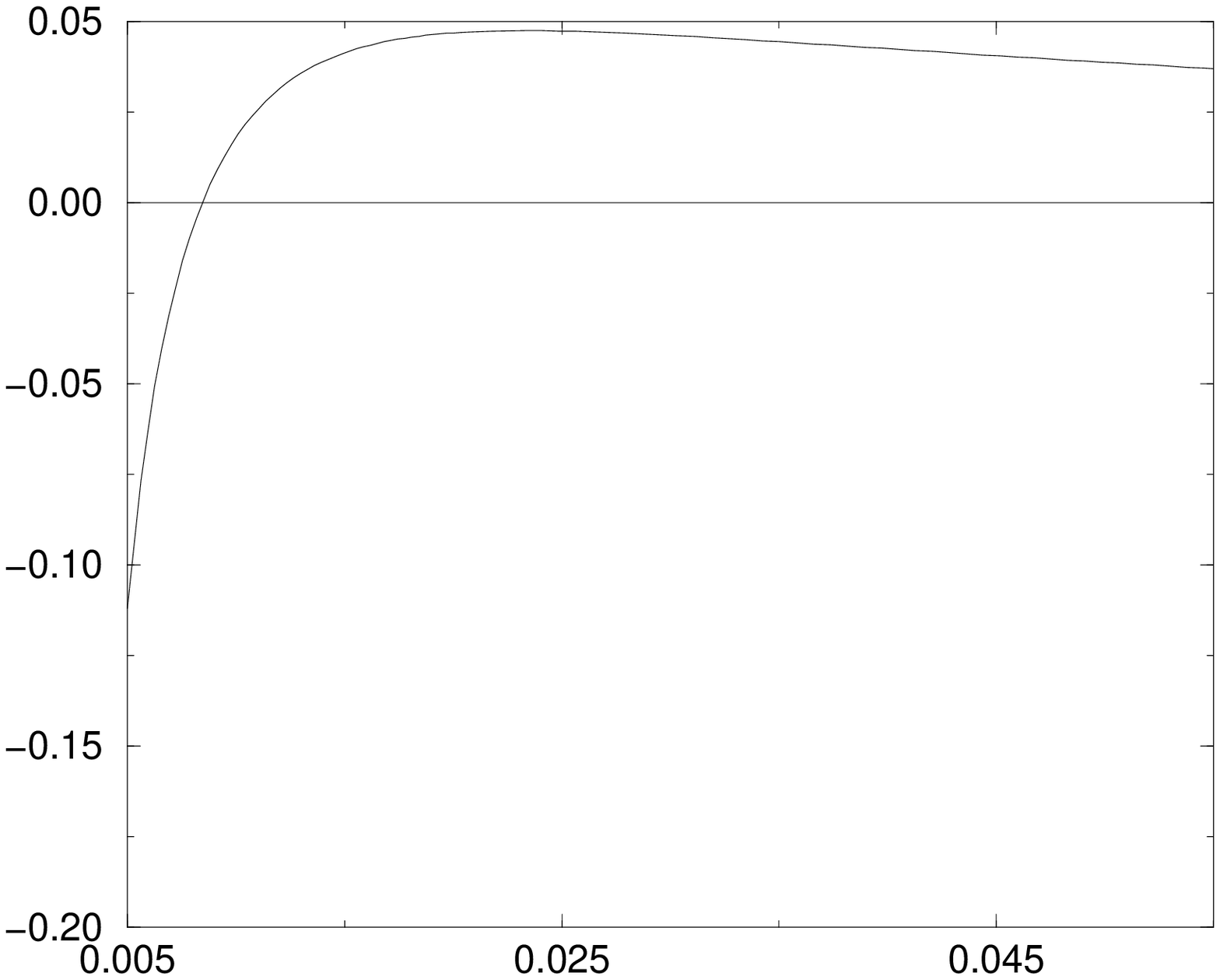}}
\put(0,0){\parbox{7cm}{Figure 2: The function $h(x)$ in case $i)$ for the
    scalar field}}
\put(5.7,1.8){$_{x}$}
\put(1.4,5.8){$_{h(x)}$}
\end{picture}
\begin{picture}(7.5,6)
\put(1,-0.8){\epsfxsize=4cm\epsfysize=9cm\epsffile{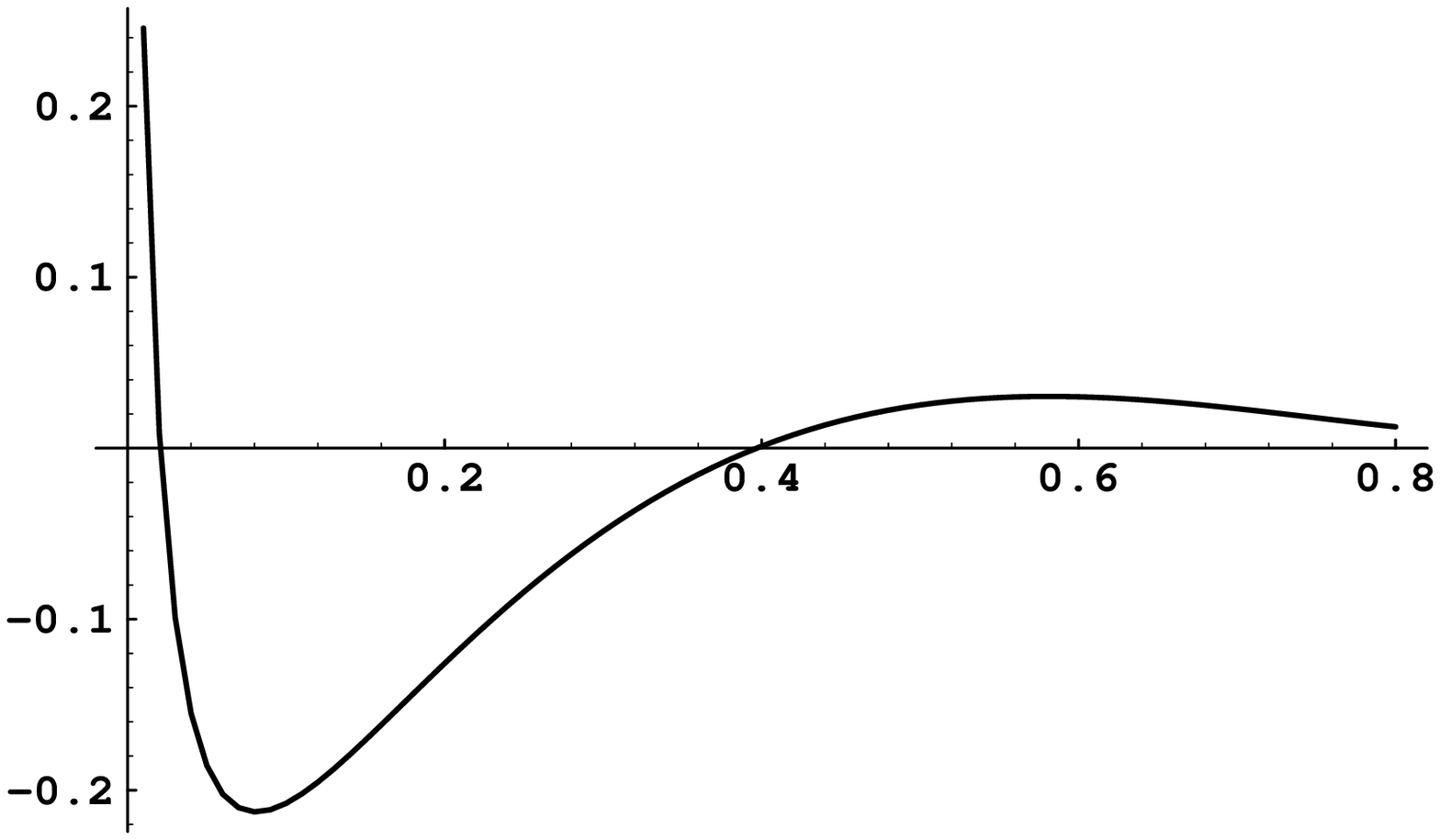}}
\put(0,0){\parbox{7cm}{Figure 3: The function $h(x)$ in case $i)$ for the
    spinor field}}
\put(1,5.8){$_{h(x)}$}
\put(5.2,3.5){$_{x}$}
\end{picture}
\label{abb3}
\end{figure}

\vspace{2cm}

\break
\noindent
The massless case deserves a special consideration. 
It was not considered in the cited papers \cite{ms,sp,gz3d}.  
It is known \cite{bvw} to exhibit the conformal anomaly which is proportional
to $a_{2}$ so that in case $a_{2}\ne 0$ the function $f(Rm)$ in \Ref{29} must
contain a divergent contribution  proportional to $\ln (RM)$ for $m\to 0$. 
Therefor, the normalization condition \Ref{normcond} cannot be extended to the
massless case. One is left with the fact that the ground state energy cannot
be uniquely defined. For example, in the model 2, i) for $m=0$ it takes in
zetafunctional regularisation the form
\[\E_{\rm qu}={-a_{2}\over 16\pi R}\left({1\over 2s}+\ln
  (R\mu)-1\right)+{\tilde{h}\over R},
\]
where $\tilde{h}$ is some number. The classical contribution must be taken as
\[
\E_{\rm class}={h\over R}
\]
in order to absorb the pol term. But a contribution of the kind $\tilde{h}/R$
resulting from the calculation of the ground state energy has the same
dependence on $R$ as $\E_{\rm class}$ and, hence, it has no predictive
power. It suffers also from the arbitrariness resulting from the parameter
$\mu$.

The situation is different in case $iii)$. There the divergencies in the
$\frac 1 R$-contribution cancel and there is no longer the need to include
$\frac h R$ into the classical energy (cf. \Ref{Eclassiii}).  The result 
\be
\tilde{h}=0.00282\label{aiii}
\ee
is unique and has a physical meaning. Therefor, of course, it coincides with
earlier calculations.

The just given considerations are of relevance for the electromagnetic Casimir
effect on a sphere. The field is massless and for the second photon
polarisation which results in Robin boundary conditions the formulas are
essentially the same (the heat kernel coefficients $a_{n}$ take different
values, see \cite{ddim} for example).

Considering an infinitely thin conducting spherical shell, the effect is
uniquely defined. If, however, one considers a spherical shell of finite
thickness with no field inside we have the cases $i)$ and $ii)$ with different
radii. No cancelation of divergencies between inside and outside
occurs. In that case it seems impossible to give a physical meaning to the
Casimir energy. 

We add just another remark concerning the electromagnetic Casimir effect for a
thin spherical shell. In that case the divergent contribution to the energy is
zero in the zeta-functional regularisation. Therefor one can obtain the finite
result without any renormalisation. In contrast, in different regularisations,
two kinds of divergencies are present, for example from \Ref{regep} we obtain
\be \E_{\rm qu}^{\rm div}={3\over 32 \pi^{3/2}}\left({a_{1/2}\over
    2\ep^{3/2}}+{a_{3/2}\over \ep^{1/2}}\right).\ee Now, $a_{3/2}$ is
independent of the radius $R$ and can be removed by arguments like saying that
only the force or the difference in the energy between two conducting spheres
of different radii has a physical meaning. The remaining coefficient
($a_{1/2}=-2\pi^{3/2}R^{2}$ for Dirichlet boundary conditions) turns out to
cancel when adding the second photon polarisation which corresponds to Robin
boundary conditions. This is the reason why it is possible to obtain a finite
result for the electromagnetic Casimir effect on a sphere in other than the
zeta-functional regularisations too. This cancelations of divergencies was
observed already in the first calculation made by Boyer. Nevertheless it is a
rather special feature and does not survive for instance when including
radiative corrections. As it was shown in \cite{rk}, the radiative corrections
to the Casimir effect for a thin conducting spherical shell yield a
divergent contribution 
\be \E_{\rm qu}^{\rm div}=-{16\over 9\pi}\alpha m_{\rm
  e}^{3}R^{2}-{4\over 15\pi}\alpha m_{\rm e}\,,\ee where $\alpha$ is the fine
structure constant and $m_{\rm e}$ is the electron mass. While the second term
could again be considered as a constant, the first did not cancel between the
two polarisations of the photon.

\vspace{2cm}

\noindent
I would like to thank the organisers of the meeting for kind
hospitality and G. Barton, K. Kirsten  and V. Mostepanenko  for
stimulating discussions.

\pagebreak

\end{document}